\documentclass[%
 reprint,%
 amssymb, amsmath,%
 aip,cha,%
]{revtex4-1}

\usepackage{docs}%
\usepackage{bm}%
\usepackage{graphicx}%
\usepackage[colorlinks=true,linkcolor=blue]{hyperref}%
\expandafter\ifx\csname package@font\endcsname\relax\else
 \expandafter\expandafter
 \expandafter\usepackage
 \expandafter\expandafter
 \expandafter{\csname package@font\endcsname}%
\fi
\begin{document}

\title{Molecular dynamics study of the linear viscoelastic shear and bulk relaxation moduli of poly(tetramethylene oxide) (PTMO)}%

\author{Zakiya Shireen}%
\email{zakiya.shireen@unimelb.edu.au}
\author{Elnaz Hajizadeh}
\affiliation{ Department of Mechanical Engineering, Faculty of Engineering and Information Technology, The University of Melbourne, Melbourne, Australia}
\author{Peter Daivis}
\affiliation{Physics Discipline, School of Science, RMIT University, Melbourne, Australia}
\author{Christian Brandl}
\affiliation{ Department of Mechanical Engineering, Faculty of Engineering and Information Technology, The University of Melbourne, Melbourne, Australia}
\date{\today}

\begin{abstract}
	Here we report the linear viscoelastic properties of amorphous poly(tetramethylene oxide) (PTMO), which is one of the key components in synthesizing segmented polyurethane (PU) elastomers. The temperature and molecular weight dependent viscoelastic behavior is investigated in detail by computing the shear relaxation modulus $G(t)$ and the bulk relaxation modulus $K(t)$, using the Green-Kubo relationship with correlation function. Our results provide new data for PTMO melt from the united atom model and also extend the existing knowledge of viscoelastic properties of polymers in general. The predicted viscoelastic behavior range is shifted on a master curve using the time-temperature superposition principle (TTSP) with horizontal and vertical shift factors. The emerging shift factors agree with the Williams-Landel-Ferry (WLF) equation. For the validation of the united-atom model of PTMO using the TraPPE-UA force field we explored the transport properties and observed a position-dependent diffusion dynamics throughout the polymer chain, which subsequently influences the scaling laws for chain dynamics. These findings are discussed in terms of emerging experimental evidence on position dependent displacement for different chain portions along the chain length.
	
\end{abstract}
\maketitle


\section{\label{sec:level1}Introduction}
	Polyurethanes (PUs) are a broad range of polymeric materials that are widely used in a variety of engineering applications, including paints, liquid coatings, insulators, adhesives, membranes \cite{wang2017facile,akindoyo2016polyurethane,rafiee2015synthesis,chattopadhyay2007structural,krol2007synthesis,petrovic1991polyurethane} etc. The significance of PUs is in the tunability of their microstructure to achieve certain desired product performances such as high mechanical strength and high toughness \cite{yan2020self,paraskevopoulou2020mechanically}, high acoustic attenuation \cite{gwon2016sound,poupart2020elaboration}, and specific optical \cite{rubner1986novel}, electrical \cite{kim2010graphene,yousefi2012self}, chemical \cite{buckley2010elasticity} and biological \cite{grasel1989properties} properties. 
	
	The microstructure of PUs is usually characterized as a two-phase composition of "soft" and "hard" segments chemically linked together along a polymer backbone. The hard segments are generally rigid and obtained by reaction of diisocyanates with diol/diamine chain extenders and include strongly hydrogen bonded urethane or urea  groups. The soft segments consist of relatively flexible polyamine/polyol chains in which the hard domains are embedded and act as a flexible continuous matrix for the network of hard domains \cite{,kojio2020influence,wei2015morphology}. The poly(tetramethylene oxide) (PTMO) is widely used as the soft segment in segmented block copoly(ether ester) PU, contributing to the microphase segregation of polyurethane elastomers \cite{prathumrat2021comparative,rahmawati2019microphase,pangon2014influence,castagna2012role}. The role of polyols as soft segments is significant in synthesising PUs because their molecular weight determines the flexibility of the PU. Considering the importance of PTMO for synthesising segmented polyurethane elastomers, it is vital to establish a link between their molecular structure and macroscopic properties. 

	Many experimentally measured properties of PTMO are widely available in literature \cite{imada1965structural,rosenberg1967mechanism,huglin1968cohesive}, including the most important static and dynamic mechanical properties \cite{mammeri2005mechanical,jewrajka2009polyisobutylene,strawhecker2013influence}. Recently, Lempesis et al. \cite{lempesis2016atomistic}, investigated the PTMO melt as a necessary precursor to model the semicrystalline PTMO system at the atomic level, using Molecular Dynamics (MD) simulations without addressing the rheological bulk behaviors. Within the processing-structure-properties paradigm, capturing the missing transport and viscoelastic behavior of PTMO is essential to design materials with enhanced and targeted properties at the atomic and molecular level \cite{hajizadeh2014nonequilibrium}. Viscoelasticity is a key characteristic of polymer melts, which carries information of the underlying relaxation processes in the systems. 

	The transport properties of polymer melts are well-captured using computational tools. Thus, the systematic numerical experiments using MD simulations are required to explore the experimentally inaccessible properties as they will help in relating the local properties to micro-structural features. Beyond the predictions of materials properties in experimentally inaccessible regimes, the predictive nature of MD provides vital data for mesoscale and continuum-scale simulation methods allowing a multiscale modelling approach directly addressing multiple length and time scales. 

	The macroscopic viscoelastic and transport properties of polymers depend on phenomena occurring at multiple length and time scales \cite{de1976dynamics,de1979scaling,doi1988theory,ferry1980viscoelastic,likhtman2002quantitative,hajizadeh2015molecular}. To obtain the full viscoelastic response of polymers by computational means, equilibrium MD and Monte Carlo simulations are used to determine the stress auto-correlation function (SACF). The SACF, when combined with the Green-Kubo relationship, allows us to calculate the viscoelastic behavior of the melt \cite{heyes1991transport,sen2005viscoelastic,zhang2018mechanical}. To probe polymer response across multiple time and length scales, various particle-based models from simple bead-spring \cite{kremer1990dynamics,hsu2016static} to coarse-grained (CG) \cite{salerno2016resolving,zhang2018mechanical}, to united-atom (UA) \cite{paul1995optimized} to highly complex all-atoms (AA) \cite{chen2008comparison} models have been developed with varying efficiency and accuracy. 
	
	The bead-spring and coarse-grained models are powerful simulation methods capturing characteristics as described in scaling theories disregarding atomistic and chemical details of the polymer chains. On the other hand the explicit all-atoms model makes it computationally expensive to probe the long-time response. To study the emergence of scaling regimes and still incorporate the most important chemical detail of the polymer, we focus on the united-atom model to study PTMO.

	The aim of this report is to enhance the understanding of viscoelastic and dynamic behaviors of PTMO melt beyond generic bead-spring models. With an explicit formation of molecular chemistry, we investigate the emergence of dynamical and viscoelastic behavior in the range from Rouse to reptative modes of PTMO model systems with accurate and site-specifically resolved chemistry within the monomer. By using a united-atom model for the PTMO melt system in equilibrium MD simulations, we establish and cross-validate the relations of viscoelastic properties to materials transport and associated structural properties (i.e., molecular weight). Fortunately, simulation and experimental studies of the time dependent shear relaxation modulus $G(t)$ for many polymeric materials are widely available. Unlike the shear relaxation modulus, studies of the time dependent bulk relaxation modulus $K(t)$ of molten polymeric systems are scarce but it is an important property in many applications, including sound attenuation. To our knowledge, there is no information available related to bulk relaxation modulus in the case of PTMO.

	This report is organized as follows. Section \ref{sec:level2} describes the model employed to simulate a PTMO melt, using the TraPPE-UA force field. In Section \ref{sec:level3}, the findings of the equilibrium MD simulations are reported in three subsections. In section \ref{sec:level3.1}, validation of the model by predicting the experimentally measured density is described. The motion of all atoms and only the inner atoms is analyzed and compared with experimental observations in section \ref{sec:level3.2}. The viscoelastic response of PTMO is discussed as shear and bulk relaxation modulus in section \ref{sec:level3.4}. Finally, section \ref{sec:level4} summarizes the conclusions emphasizing how the results from this work can be used in further research.

\section{\label{sec:level2}Method}

	The repeat unit $-(CH_2CH_2OCH_2CH_2)-$ for the PTMO chain was generated using the molecule builder moltemplate \cite{jewett2013moltemplate}, where every oxygen and methylene group was treated as an atom and united atom \cite{paul1995optimized}, respectively. Random paths were generated using a self-avoiding random walk for long polymers chains. A Monte Carlo procedure is used to generate multiple paths, while ensuring unbiased sampling of the lattice Hamiltonian path \cite{mansfield2006unbiased}. These paths do not resemble the natural molecular curvature of a polymer chain as each step in the path corresponds to a location in the lattice. Therefore, subsequently, the polymer chains are relaxed using the united-atom potential allowing a more natural curvature of the molecules.

	The united-atom potential used in this work is the Transferable Potentials for Phase Equilibria-UA (TraPPE-UA) \cite{martin1998transferable,wick2000transferable} for the intra and inter-molecular bonded and non-bonded interactions. The potential forms and force field parameters are adopted from the work of Lempesis et al. \cite{lempesis2016atomistic}, where a harmonic bond-stretch potential was chosen instead of the original TraPPE-UA with stiff bond stretching potential. The parameters for the harmonic bond stretching potential term were taken from the Optimized Potentials for Liquid Simulations-UA (OPLS) \cite{jorgensen1984optimized,jorgensen1988opls} force field and they are listed in Table S1 (Supplementary information). Non-bonded interactions in the TraPPE-UA potential include pairwise Lennard-Jones $(LJ)$ and Coulombic potentials. The Lorentz-Berthelot combining rules were used to determine the parameters for unlike $LJ$ interactions. Non-bonded interactions were calculated only for inter and intra-molecular interactions of atoms separated by four or more bonds. The long range Coulombic interactions were calculated via the particle mesh Ewald method \cite{darden1993particle,essmann1995smooth} and intra-molecular interactions separated by three bonds were reduced in magnitude by a scaling factor of 0.5 \cite{stubbs2004transferable}. To achieve correct liquid-state thermodynamic properties as outlined in a previous study \cite{huang2010effect}, the cut-off value is $\approx 3\sigma$. Therefore, in this study, both $LJ$ and the short ranged part of the electrostatic interactions are truncated at a cut off value of $11$ \AA. 

	We performed molecular dynamics (MD) simulations, using the LAMMPS software package \cite{plimpton1995fast} to simulate an initial PTMO melt configuration with the experimentally measured density \cite{tsujita1973thermodynamic} of $\rho = 0.87$ $g/cm^3$. The simulation box consisted of $20$ chains of equal length i.e., $N_m = 555$ repeat units (monomers). Thus, every polymer chain has total of $2775$ sites, which corresponds to a molecular weight of $M = 39960$ g/mol and the total number of atoms in the system is $55,500$. The equations of motion were integrated using the velocity-Verlet method with a time step $1$ fs. The system was equilibrated at $T = 453$ K for $5$ ns in a canonical ensemble using the deterministic Nose-Hoover thermostat with time constant equal to $0.1$ ps \cite{nose1984unified,hoover1985canonical}. This temperature is above the previously reported melting temperature of PTMO at atmospheric pressure, i.e., $T_m = 316$ K \cite{imada1965structural,huglin1968cohesive,trick1967crystallization}. 

	Subsequently, the equilibrated system was quenched in the $NPT$ ensemble at $P = 1$ atm, from $453$ K  to $313$ K in decrements of $20$ K. While quenching, isothermal and isobaric conditions were maintained with constants $0.1$ and $1$ ps, where at each temperature the simulation ran for $5$ ns. Note that, the system with $N_m = 555$ is a reference system to validate our united-atom model by predicting densities and comparing with the experimental density value at specific temperature and pressure. 
	
	Considering the limitations of MD in terms of achievable time scales within a reasonable computational time frame, we have considered smaller chain lengths i.e $N_m = 50$, $90$ and $200$ at elevated temperatures for dynamic computations. 
	
\section{\label{sec:level3}Results and discussion}
\subsection{\label{sec:level3.1}Specific volume}
\begin{figure}
	\includegraphics[width=0.5\textwidth]{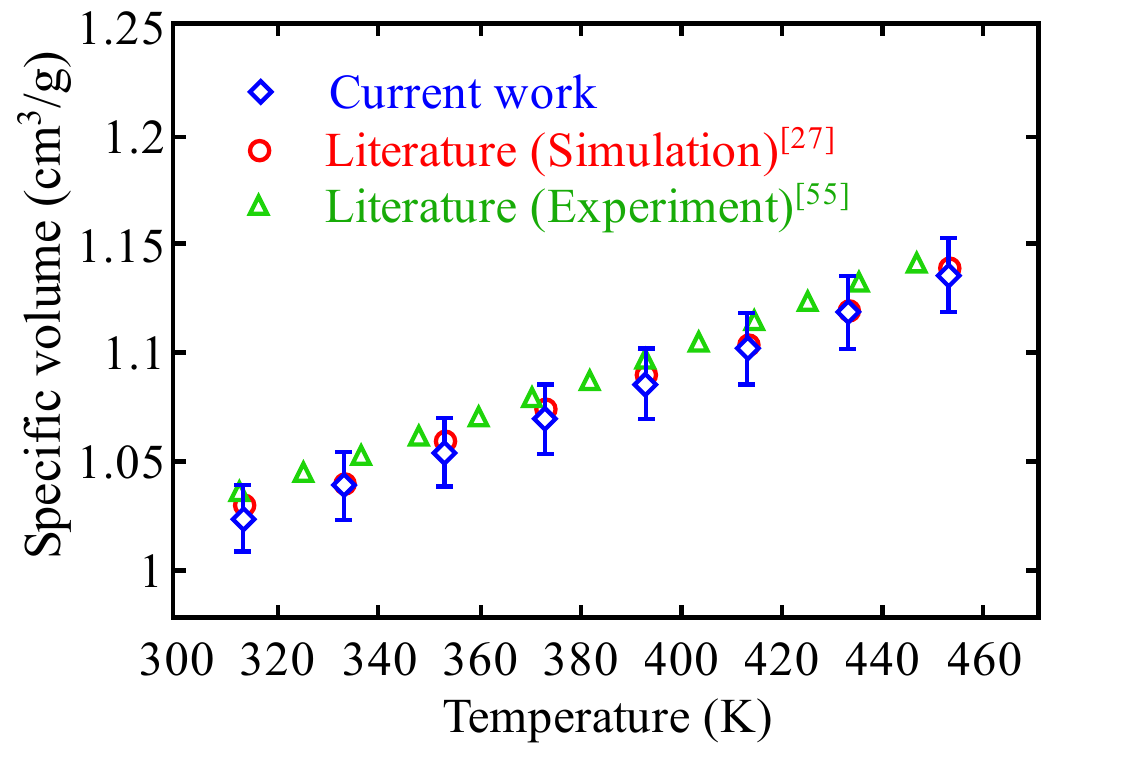}
	\caption{Comparison between simulated volumetric behavior of a PTMO chain with $N_m = 555$ repeat units (diamonds) and other simulation values (circles) \cite{lempesis2016atomistic} as well as experimental measurements (triangles) \cite{tsujita1973thermodynamic}.}
	\label{Figure:1}
\end{figure}
	Before conducting the united-atom simulations to study the viscoelastic properties of PTMO the simulation approach and interatomic potential were validated by calculating the temperature-dependent specific volume at $P = 1$ atm. Fig. \ref{Figure:1} shows the specific volume of a PTMO melt as a function of temperature as well as experimental and previously reported molecular simulation results \cite{lempesis2016atomistic,tsujita1973thermodynamic}. The maximum deviation of the simulation results obtained in the present work is observed to be less than 1\% for all cases. To study the molecular weight dependency of thermodynamic properties, systems with different molecular weights (i.e., different degree of polymerization) were simulated. 

	The subsequent studies focus on computationally inexpensive smaller chain lengths i.e., $N_m = 50$, $90$ and $200$ to investigate the dynamics and viscoelastic behavior of PTMO, allowing us to access the required time scale to observe the underlying relaxation processes. It is important to note that we observed the deviation of less than 2\% for densities at smaller chain lengths for a given temperature and pressure.	
	
	\subsection{\label{sec:level3.2}Finite Size Effect}
\begin{figure}
	\includegraphics[width=0.5\textwidth]{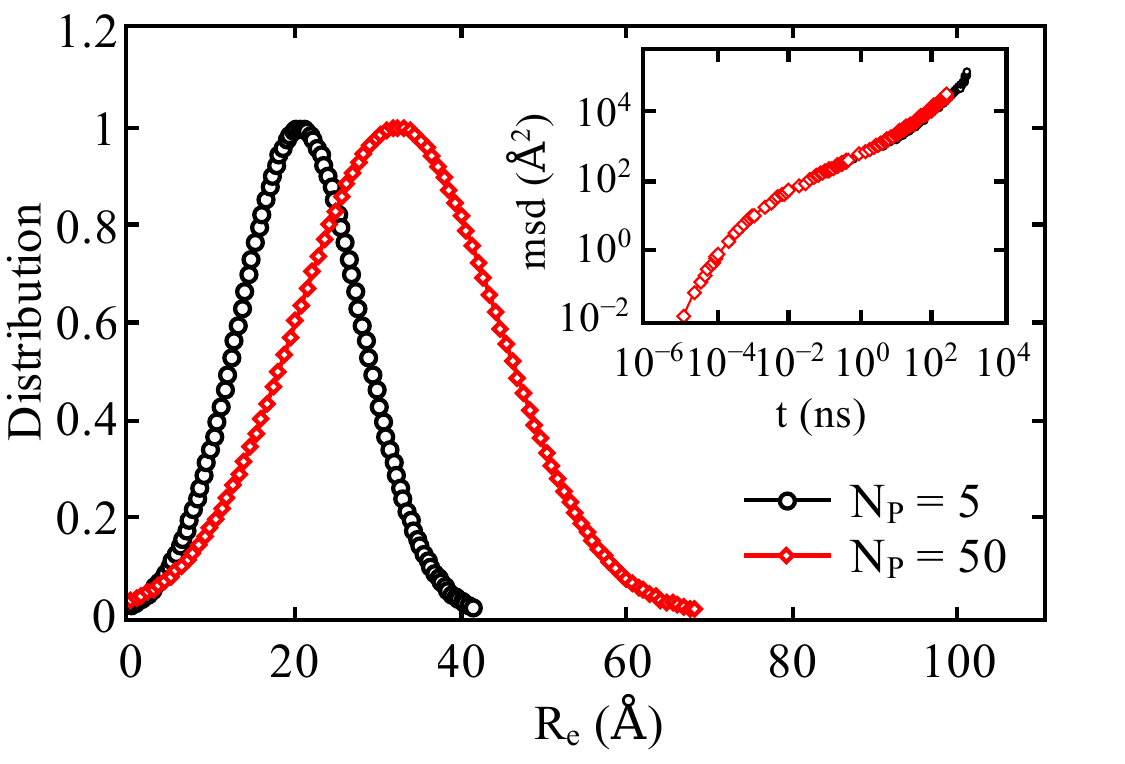}
	\caption{Distribution of root-mean-squared end-to-end distance ($R_e$) is shown for $N_P = 5$ and $50$. In the inset mean-squared-displacement (MSD) observed from two systems are shown for chain length $N_m = 50$.}
	\label{Figure:2}
\end{figure}
	Polymer theory holds true when the system size tends to infinity. With computational limitations in simulations we always work with finite box sizes,which might influence the results. In order to predict accurate long time behavior of PTMO melts we compared two system sizes by varying number of polymers ($N_P$) and associated box sizes for a given chain length. We present results on the root-mean-square end-to-end distance $\langle{R_e}^2\rangle^{1/2}$ distribution from a system with $N_P = 5$ and $50$ in Fig. \ref{Figure:2}. The simulations were performed for the chain length $N_m = 50$ and the associated simulation box size for the two systems are $33.07$ and $71.6$ {\AA}, respectively. With increasing $N_P$ we observe wide end-to-end distributions and the averaged root-mean-squared value of end-to-end distance $R_e$ is $21.70$  {\AA} and $34.47$  {\AA} for $N_P = 5$ and $50$, respectively. Using the freely-jointed chain model ($R_e = n{l_b}^2$), the estimated value of end-to-end distance for $N_m = 50$ is $23.62$  {\AA} where $l_b$ is root-mean-square bond length $\langle{l_b}^2\rangle^{1/2}$. Here the root-mean-square bond length $\langle{l_b}^2\rangle^{1/2}\approx1.497$ {\AA}. The computed $R_e$ for $N_P = 50$ is larger than theoretically estimated (Freely-jointed-chain model) value due to shorter chain length and semi-flexible bonds. 
	
	We also tested the effect of system size on the dynamics and mean-squared-displacement (MSD) for two system sizes and is shown in the inset of Fig. \ref{Figure:2}. Diffusion of PTMO chains are discussed in detail in section \ref{sec:level3.3}. The similar test was done for stress relaxation (section \ref{sec:level3.4}) and we observe that the results are not affected by the system size although the end-to-end distance shows finite size effects. In order to reach correlation function lag times up-to $t\approx10^3$ ns for long time-scale behavior of PTMO melt, system size $N_P = 5$ is considered in conjunction with ensemble averaging over multiple independent configurations. This approach allow us to use a direct MD simulations approach with readily available computer resources. The results presented here are insensitive to the finite system size unless stated otherwise.

\subsection{\label{sec:level3.3}Diffusion of the polymer chain}
	To resolve the rheological properties of PTMO melt, the transport behavior of the melt system within the MD timescales, is investigated. We study dynamic properties of polymer chains with the molecular weights $M = 3600$, $6480$ and $14400$ g/mol, which correspond to number of monomers $N_m = 50$, $90$, $200$ in one chain. The transport of polymers in the liquid state is usually characterized by the mean-squared-displacement (MSD) of the monomers. To examine the mobility of all atoms of PTMO, the MSD is computed through Eq. \ref{eq1}, where, $g_1(t)$ is the MSD of all atoms in the chain, including the chain ends, where $n_a$ is the number of atoms.

\begin{equation}
	{g_1(t)} = \frac{1}{{n_a}}\sum_{i=1}^{n_a}\langle[r_i(t)-r_i(0)]^2\rangle
	\label{eq1}
\end{equation}

\begin{figure}
	\includegraphics[width=0.5\textwidth]{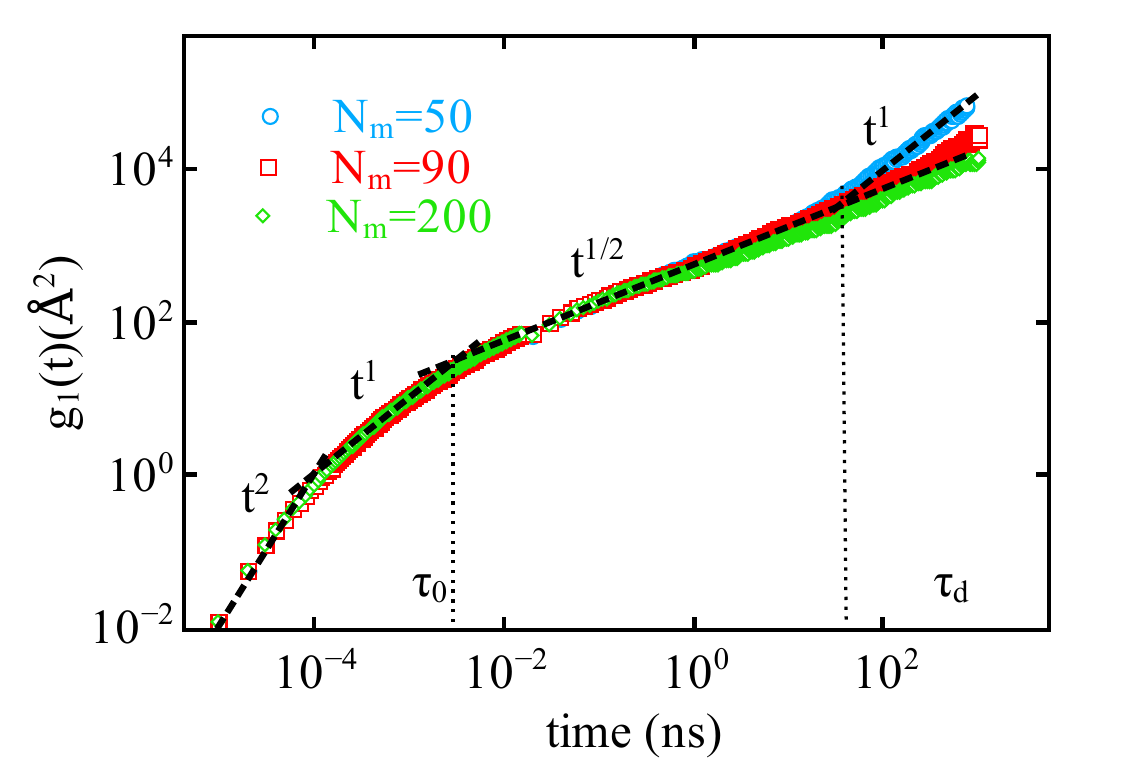}
	\caption{Mean-squared-displacement (MSD) $g_1(t)$ of all atoms as a function of time for $N_m = 50$ (circles), $90$ (squares) and $200$ (diamonds) at $T = 553$ K. The dashed lines show different scaling regimes.}
	\label{Figure:3}
\end{figure}

\begin{figure}
	\includegraphics[width=0.5\textwidth]{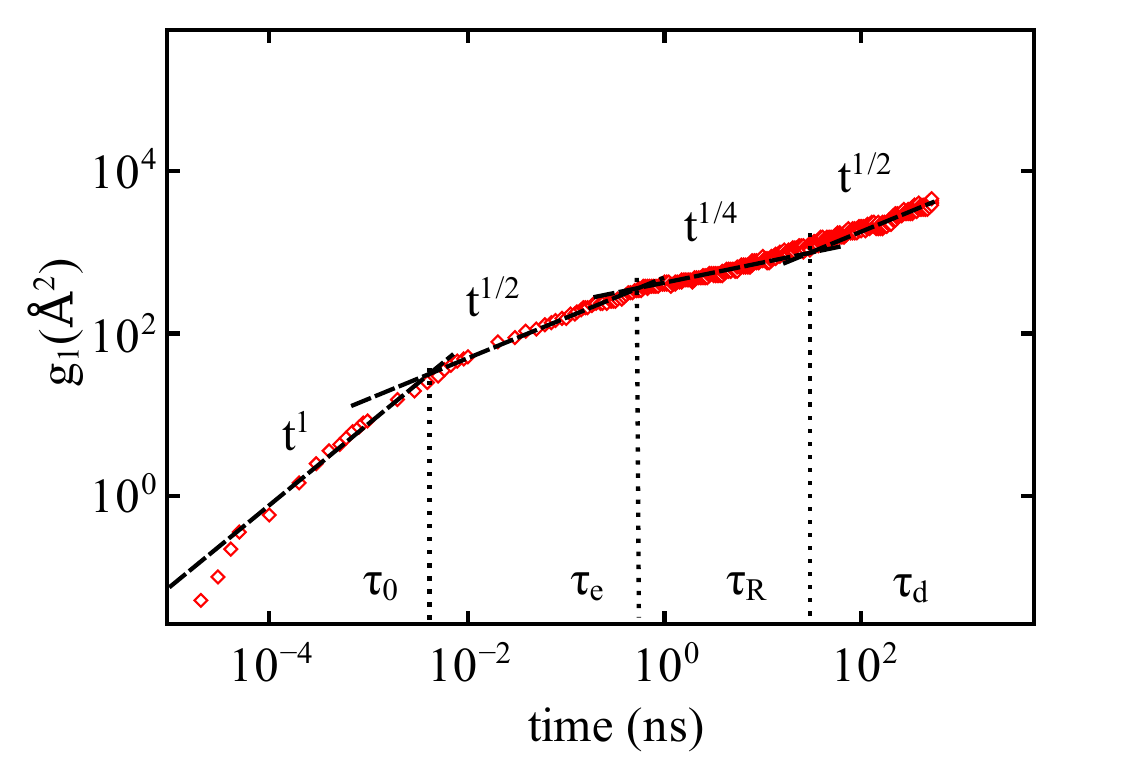}
	\caption{(a) The MSD $g_1(t)$ of middle portion of the chain i.e. 25 atoms for $N_m = 200$ at $T = 553$ K. The crossover points between the adjacent scaling regimes are determined by intersections of the fitted dashed lines.}
	\label{Figure:4}
\end{figure}

	The inherent complex topological constraints in polymer chains play an important role in dictating the dynamical properties of these materials at multiple time and length scales. To determine the dynamic scaling behavior of PTMO chains, we compare the MSDs of three different chain lengths and plot $g_1(t)$ as a function of time in Fig. \ref{Figure:3}. To probe the position dependent diffusion along the polymer chain, we consider two case: (a) $g_1(t)$ of all atoms and united atoms (UAs) in the chain (Fig. \ref{Figure:3}) and (b) $g_1(t)$ of only the inner 25 atoms and UAs in the chain (Fig. \ref{Figure:4}).

	Four distinct scaling regimes, a characteristic time scale $\tau_0$ followed by the entanglement time $\tau_e$, the Rouse time $\tau_R$, terminated by the disentanglement time $\tau_d$ characterizes the local motion of very long polymer chains \cite{hsu2016static}. To determine the agreement with theoretically predicted scaling laws, relaxation times are calculated by identifying the intersection points of the two fitted lines in the adjacent scaling regimes. Each intersection point corresponds to a crossover point between the two timescales. Fig. \ref{Figure:3} represents the averaged motion of all atoms (oxygen) and united atoms (methylene groups) in the chains, including the chain ends. For the averaged $g_1(t)$ at short timescales, we observe the higher displacements of atoms, which is given as the power law $t^2$ of the fitted line for all the three chain lengths shown in Fig. \ref{Figure:3}. It is found that the crossover from $t^2$ to early diffusive regime occurs at $t\approx 10^{-4}$ ns, which is approximately a decade on time scale and is $\ll\tau_0$. We suggest that, the higher displacement is a result of enhanced motions of atoms at the chain ends.

	Recently a similar experimental finding was reported in the study of double-stranded DNA (dsDNA) molecules, where super-resolution (SR) fluorescence localization microscopy and cumulative-area (CA) tracking were used to track a single molecule \cite{abadi2018entangled}. In the analysis of experimental observations, displacement obtained between the contours for adjacent time points revealed that the local chain displacement along the direction perpendicular to the contour increases towards the chain ends and displays displacement larger than the tube diameter at the chain ends \cite{abadi2018entangled}. The similar observation of "large motional freedom" at chain ends has also been observed in synthetic polymers \cite{keshavarz2016nanoscale,keshavarz2017confining}. The growth of the MSD is eventually slowed down by the less mobile atoms at the middle of the chains.

	For polymer diffusion, it is expected that positions (outer and inner) of atoms along the chains strongly influence its diffusion behavior. In Fig. \ref{Figure:3}, the crossover time $\tau_0$ appears to be the same for all $N_m$. As time increases, it is observed that the crossover to $t^{1/2}$ occurs at $t\approx 10^{-2}$ ns, which is an indication of the Rouse relaxation \cite{kremer1990dynamics}. For $N_m = 50$, we observe a direct transition of this $t^{1/2}$ scaling regime to the free diffusion regime at $t\approx 10$ ns. The polymer chain with $N_m = 50$ is likely to be in an unentangled state and is found to behave in accordance with the Rouse theory. However, as the chain length increases the $t^{1/2}$ scaling persists beyond $t = 100$ ns and for $N_m > 50$ the onset of slowing down occurs at the same time ($\approx 10^{-2}$ ns).

	According to the reptation theory predictions, the Rouse-like behavior i.e $t^{1/2}$ in $g_1(t)$ should become a $t^{1/4}$ power law for chain lengths above the entanglement length \cite{de1979scaling,doi1988theory}. The MSD averaged over all atoms and united atoms in the chain including chain ends does not reveal the crossover from Rouse to reptation regime as predicted by the tube model. This can possibly be explained on the basis of a recent experimental observation, where for chain position-dependent displacement a timescale of the order of minutes, shorter than the Rouse time $\tau_R$ ($2.3- 3.3$ hours) is observed for dsDNA molecule. This experimental observation implies that the chain position-dependent motion is not directly related to the reptation motion of the chain. The authors suggest that this timescale is an unambiguous sign of the constraint release motion of polymer chains that are captured at subchain-level of entangled polymer in real space \cite{abadi2018entangled}.

	In the present study the position-dependent displacement due to the constraint release mechanism \cite{klein1986dynamics,graessley1982entangled,kavassalis1988new}, is not identified explicitly but the scaling of the Rouse regime is extended for up-to approximately $5$ decades (from $\approx 10^{-2}$ to $10^{3}$ ns) for $N_m = 200$. In the study of synthetic polymer melts, it has been suggested that the constraint release motion influences self diffusion of an entangled polymer when the molecular weight is in the range between $M_e < M < 10M_e$ \cite{von1991reptation,smith1985polymer} and in our system the entanglement length $N_e$ is estimated to be $21.4 \pm 2.0$ which corresponds to entanglement molecular weight $M_e = 1396.8 - 1684.8$ g/mol for $N_m = 200$ . Therefore, the local chain displacements at intermediate timescale (identified as $t^{1/2}$ from intermediate to long timescale) for $N_m = 200$ ($M = 14400$ g/mol) can probably be interpreted by the constraint release motion of the chain but further investigation is required to quantify this.

	Fig. \ref{Figure:4} visualizes the diffusion of the inner atoms from the middle segment of the chains and present $g_1(t)$ averaged over $25$ atoms for $N_m = 200$. The MSD averaged only over $25$ inner atoms displays the apparent power law behavior as predicted by reptation theory, from the intermediate to long timescales. Therefore, to observe the position-dependent-displacement of entangled polymer chains (consistent with experiments) and identification of the exact scaling regimes at subchain-levels requires that the multiple segmental dynamics is probed.

\subsection{\label{sec:level3.4}Viscoelastic Properties}
	The viscoelastic behavior of polymers is usually solely characterized by the shear relaxation modulus $G(t)$ which, we here extend to the bulk relaxation modulus $K(t)$ \cite{ferry1980viscoelastic}. We also explored the temperature and molecular weight dependence of the relaxation moduli. 

	To capture the viscoelastic properties of PTMO melts, we employ the multi-tau correlator method \cite{ramirez2010efficient} to calculate the stress auto-correlation function (SACF) in a canonical ensemble (NVT). Moreover, we increase the simulation time up-to $10^2-10^3$ ns , which provides sufficiently accurate estimates of the linear rheological properties of PTMO melts. Time-dependent shear relaxation modulus $G(t)$ with tensorial stress autocorrelation function (SACF) $C_{ij}$, where $i,j=(x,y,z)$ is given as:	
\begin{equation}
	C_{ij}(t) = \frac{V}{{k_B}T}\langle P_{ij}(t)(P_{ij}(0)\rangle
	\label{eq3}
\end{equation}	
	where, $V$ is the system volume, $T$ is the temperature, $k_B$ is the Boltzmann constant and $P_{ij}$ are the stress tensor components. The shear relaxation modulus $G(t)=(C_{xy}(t) + C_{xz}(t) + C_{yz}(t))/3$ is computed from the shear components in $C_{ij}$ which is justified in an isotropic fluid. 
	
	The time-dependent isothermal bulk relaxation modulus $K(t)$ is calculated from the stress deviation auto-correlation function in a canonical ensemble (NVT) given as,	
\begin{equation}
	X_{ii}(t) = \frac{V}{{k_B}T}\langle (P_{ii}(t)-\langle P \rangle )(P_{ii}(0)-\langle P \rangle )\rangle
	\label{eq4}
\end{equation}	
	where, $P_{ii}$ represents the normal components of the stress tensor and $\langle P \rangle$ is the pressure of the system (ensemble) averaged over time. The isothermal bulk relaxation modulus is computed as $K(t)=(X_{xx}(t) + X_{yy}(t) + X_{zz}(t))/3$. In this study the ${\langle P \rangle}^2$ term is taken from the long-time value of $\langle P_{ii}(t)P_{ii}(0)\rangle$ averaged over the last $50$ correlation delay times in all cases. Note that $K(t)$ defined here only represents the time-dependent part of the bulk relaxation modulus. The full relaxation modulus is the sum of this time dependent function and a constant term which can be determined from the mechanical equation of state \cite{ferry1980viscoelastic}. Given that the multiple-tau correlator method is based on the idea of pre-averaging the data, with changing averaging time, before the calculation of time correlations, \cite{ramirez2010efficient} the stresses were analyzed  every $1$ ps for smoother statistics. 

\subsubsection{\label{sec:level3.4.1}Molecular Weight Dependence}
\begin{figure}
	\includegraphics[width=0.5\textwidth]{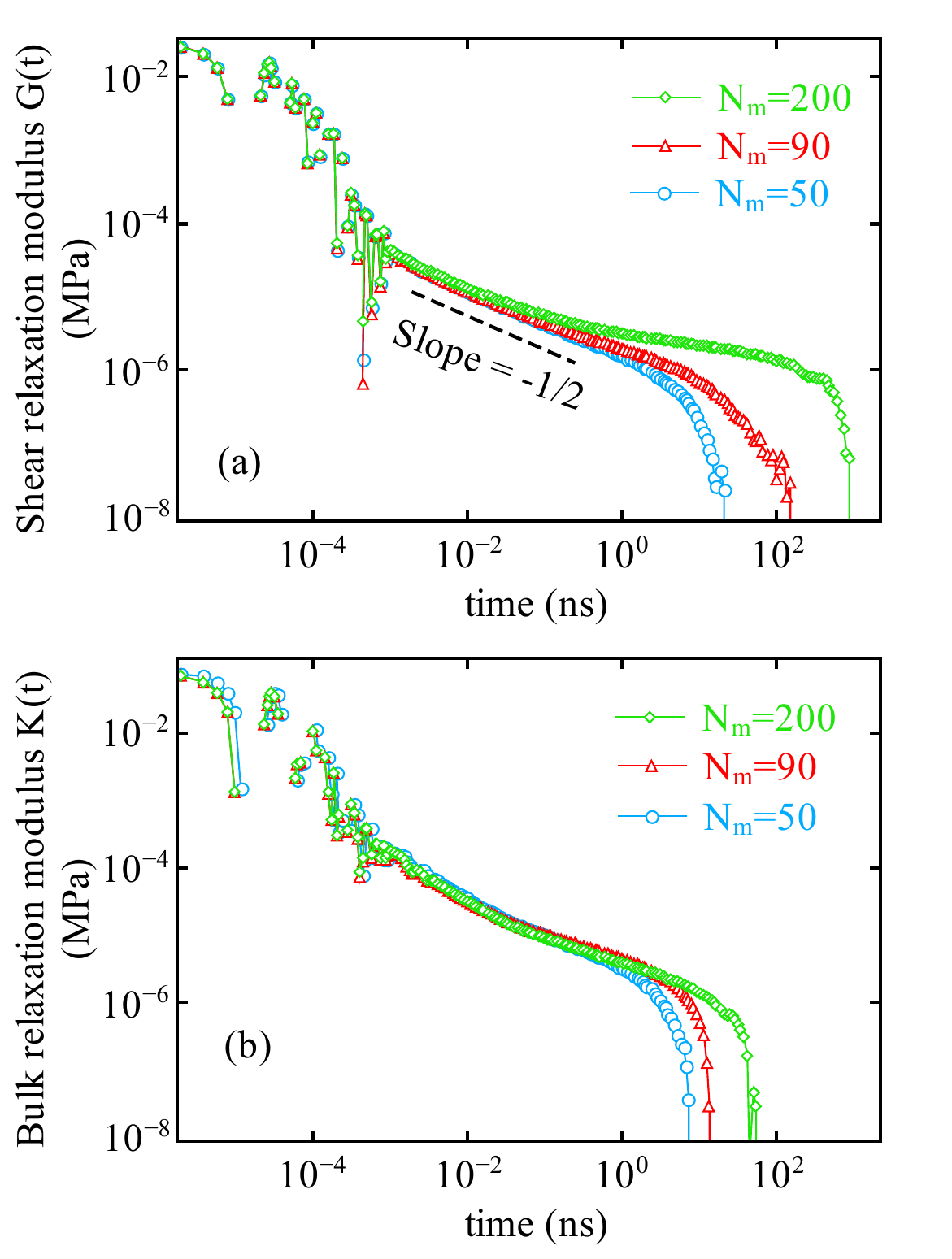}
	\caption{(a) Shear relaxation modulus $G(t)$ and (b) Bulk relaxation modulus $K(t)$ are shown for $N_m = 50$ (circles), $90$ (triangles) and $200$ (diamonds) at $T = 553$ K.}
	\label{Figure:5}
\end{figure}

	In Fig. \ref{Figure:5}(a,b) the shear relaxation modulus $G(t)$ and bulk relaxation modulus $K(t)$ respectively, are plotted as a function of time for $N_m = 50$, $90$ and $200$ at $T = 553$ K. The Rouse behavior \cite{rouse1953theory} $G(t)\sim t^{-1/2}$ is observed and shown by the dashed line in Fig. \ref{Figure:5}a. For $N_m = 50$, we only observe the Rouse relaxation as a result of unentangled short chains which is consistent with observed diffusion in Fig. \ref{Figure:3}.

	Theoretically, the relaxation of polymer chains is characterized by the Rouse behavior at $t<\tau_e$, while $G(t)$ reaches a plateau value depending upon the molecular weight \cite{hsu2016static}.  For $N_m > 50$ and with increasing time, we observe that $G(t)$ reaches an expected plateau value that persists for a longer time for longer chain lengths, as predicted by reptation theory \cite{de1979scaling,doi1988theory}. The plateau signifies that the polymeric chains are entangled and can not pass through each other. As a result, from $\tau_e < t <\tau_d$ polymer chains are believed to be moving in a tube-like region due to the entanglements and finally at $t > \tau_d$ chains are completely relaxed and $G(t)$ deviates from the plateau. 
	
	Since the disentanglement time scales as $\tau_d\sim N^{3.4}$, the plateau value or the constant $G(t)$ value extends to this range with increasing $N_m$ and relaxes completely at $\approx 1$ $\mu$s for $N_m = 200$. For $N_m = 200$, $\tau_e$ is estimated as $\approx 0.592 \pm 0.002$ ns from $g_1(t)$ (Fig. \ref{Figure:4}) and $\approx 0.588 \pm 0.002$ ns from $G(t)$. Using the standard expression for the plateau modulus $G^{0}_N = (4/5)(\rho k_B T/N_e)$, the entanglement length $N_e = 21.4 \pm 2.0$ is estimated for $N_m = 200$. From bead-spring model of semi-flexible polymeric chains, for $1000$ bead chain (equivalent to $N_m = 200$ of present study) with statistical segment length $0.964$ the $N_e$ is estimated $26 \pm 3$ \cite{hsu2016static}. Moreover, the $N_m = 90$ case in Fig. \ref{Figure:5}a shows no plateau and signatures of entanglement. The emerging plateau at $N_m = 200$ confirms that the $N_m = 200$ system is in transition from unentangled towards an entangled state. Of particular importance is the observation that at long times, $G(t)$ falls sharply for all chain lengths, which indicates that stress auto-correlation functions are long enough to capture the entire relaxation process.

	In Fig. \ref{Figure:5}b the bulk relaxation modulus $K(t)$ for polymer systems with three different chain lengths is plotted as a function of time. Similar to the shear relaxation modulus, we observe the re-adjustment of bonds i.e. bond relaxation at time $10^{-5}-10^{-3}$ ns for the bulk relaxation modulus. From intermediate to long time scales, we observe that the bulk relaxation modulus decays faster and as a result relaxation time is smaller ($\approx 10-10^2$ ns) for $K(t)$ than for $G(t)$ for all molecular weights. The bulk relaxation time increases with increasing chain length.

\subsubsection{\label{sec:level3.4.2}Effect of Temperature}

\begin{figure}
	\includegraphics[width=0.5\textwidth]{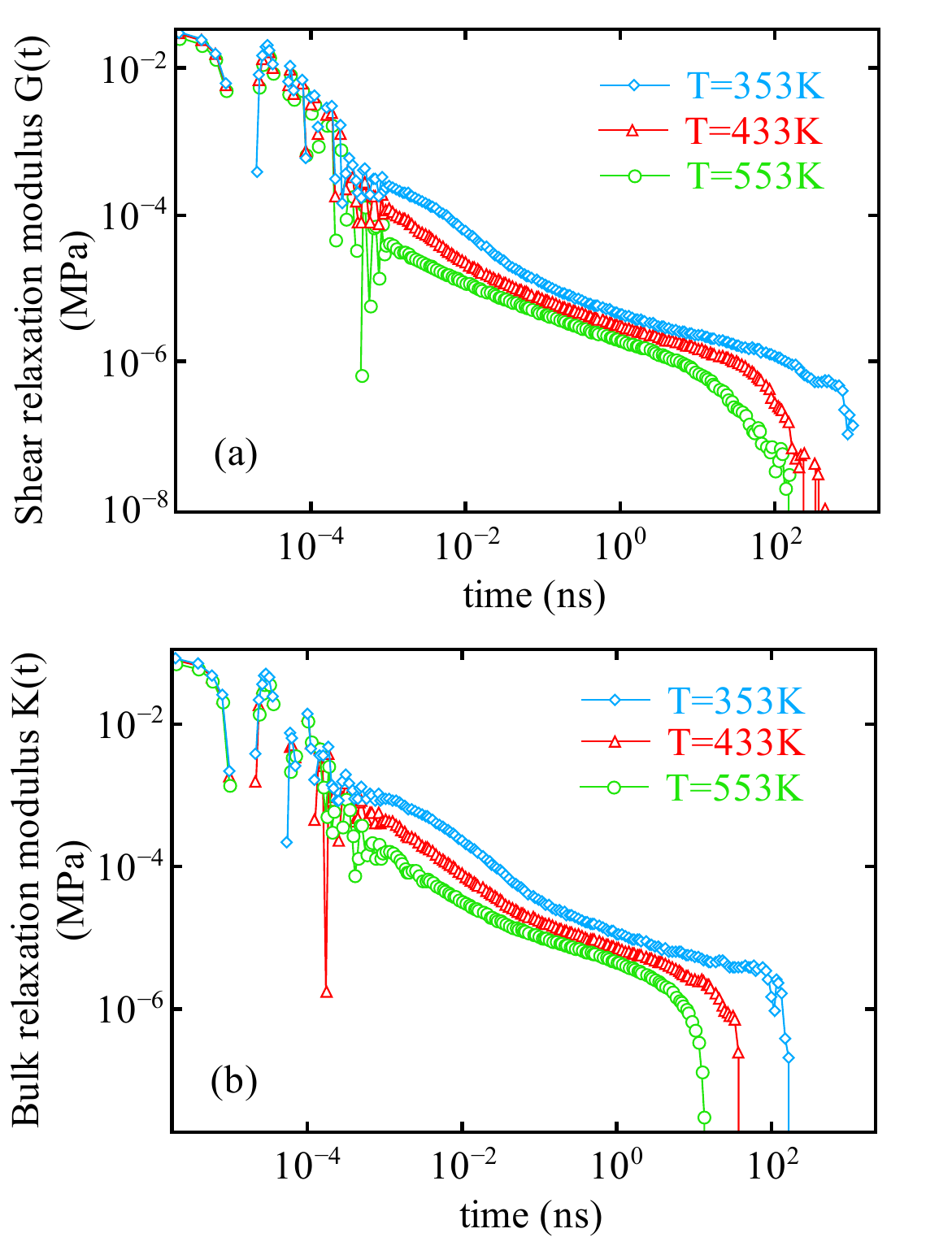}
	\caption{(a) The shear relaxation modulus $G(t)$ as a function of time for $N_m = 90$  at temperatures $553$ K(circles), $433$ K (triangles) and $353$ K (diamonds). (b) Corresponding $K(t)$ as a function of time.}
	\label{Figure:6}
\end{figure}
	In Fig. \ref{Figure:6}a, the stress relaxation of PTMO melt as a function of time is plotted for a range of temperatures at $N_m = 90$. The relaxation behavior at different temperatures is again characterized by the discussed stages signifying the melt state. At short times i.e from $10^{-5}$ to $10^{-3}$ ns, the stress relaxation occurs predominantly by the rearrangement of bond lengths i.e. bond relaxation and is independent of the temperature and $G(t)$ and $K(t)$ collapse on top of each other at short time scale. At intermediate time scale, the relaxation is the result of rearrangement of the segments of the polymer chains, with evolution of relaxation moduli at different temperatures as shown in Fig. \ref{Figure:6}(a,b). 

	With decreasing temperature the relaxation pattern remains the same but shifts along the time scale as well as vertically. The shift along y-axis is due to thermal expansion (also see Fig. \ref{Figure:1}). Delay in relaxation (i.e., shift along time scale) is due to the slow dynamics in the system. 

	Diffusion slow down is explained by the decrease of the free volume with decreasing temperature, which is linked to the thermal expansion in Fig. \ref{Figure:1}. These findings highlight that the small amount of change in free volume can have significant effect on the time dependence of the materials response. Shear and bulk relaxation modulus both show the similar temperature dependency. Therefore, we compute the shift factors to construct master curves for the viscoelastic properties.

\begin{figure}
	\includegraphics[width=0.5\textwidth]{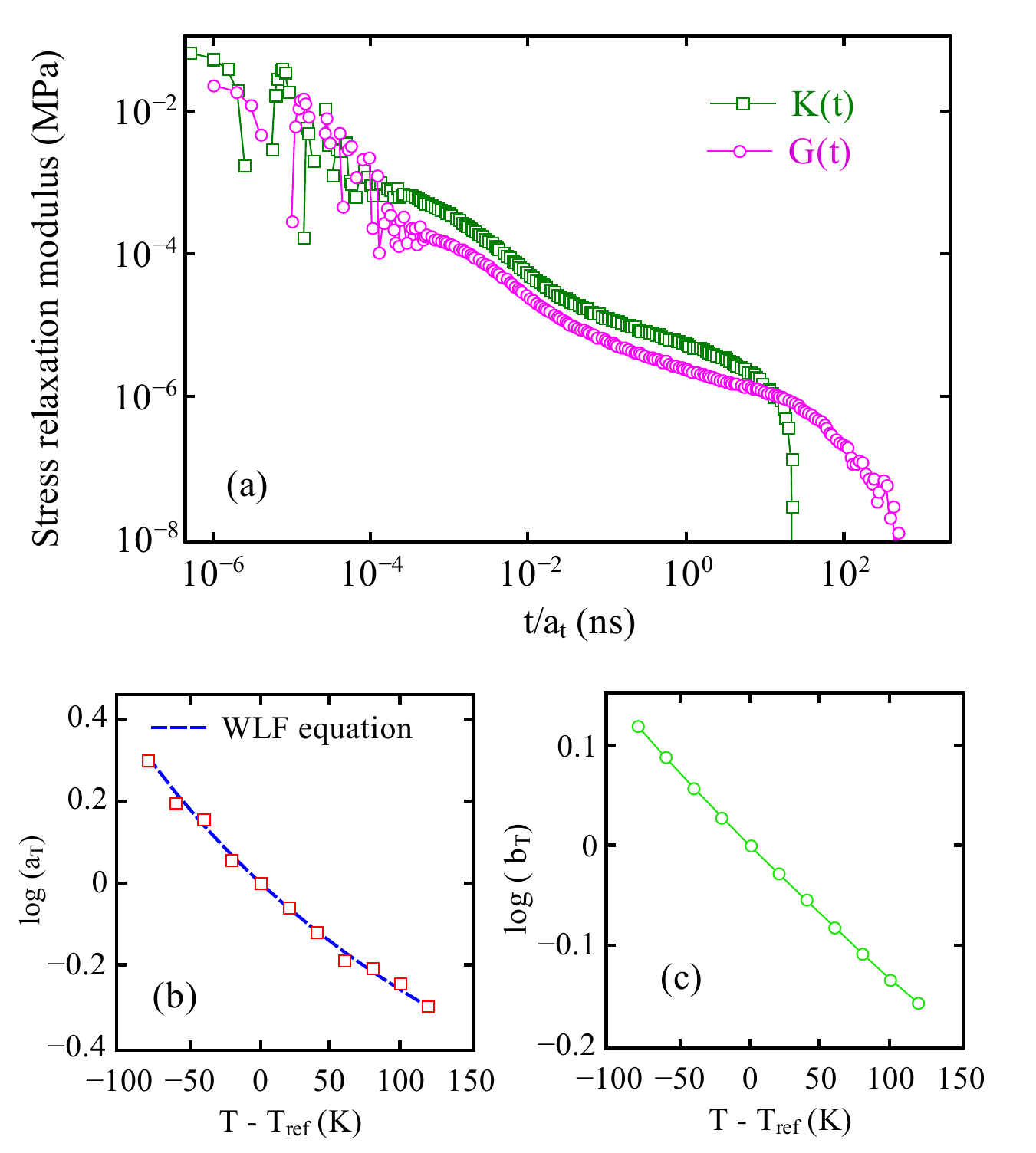}
	\caption{(a) Master curves are constructed for shear $G(t)$ and bulk relaxation modulus $K(t)$ at $T = 433$ K for $N_m = 90$. (b)The horizontal shift factors are plotted against $T-T_{ref}$ and fitted Williams-Landel-Ferry equation is shown as dashed line. The vertical shift factor $b_T$ is shown in (c).}
	\label{Figure:7}
\end{figure}
	The time-temperature superposition (TTS) principle can be applied to time-dependent stress relaxation curves calculated at different temperatures to obtain the master curve of the dynamical behavior of the system. In this approach, data obtained at one temperature are transformed to another temperature by a simple multiplicative transformation of the time scale \cite{tobolsky1956stress,rouleau2013application}. TTS is often employed to extend the values of the $G(t)$ computed at any temperature to both short and long times that can be obtained experimentally. In Fig. \ref{Figure:7}a, two master curves are constructed at a reference temperature $T_{ref} = 433$ K based on $G(t)$ and $K(t)$ data. In order to achieve the superposition of viscoelastic data, horizontal ($a_T$) and vertical ($b_T$) shift factors are computed. The $a_T$ is computed from zero shear viscosity and plotted as a function of $T-T_{ref}$ in Fig. \ref{Figure:7}b. The horizontal shift factors are fitted by the Williams-Landel-Ferry (WLF) \cite{williams1955temperature} equation given as:

\begin{equation}
	\log (a_{T}) = \frac{-C_1(T-T_{ref})} {C_2+T-T_{ref}}
	\label{eq5} 
\end{equation}

	Eq. \ref{eq5} is shown as a dashed line in Fig. \ref{Figure:7}b with parameters $C_1 = 1.47$ and $C_2 = 463$. The vertical shift factor $b_T$ reflects the temperature dependence of the moduli of a viscoelastic material and can be calculated from $b_T = {T_{ref}}\rho_{ref}/T\rho$, where $\rho_{ref}$ represents the density of the PTMO at reference the temperature \cite{dealy2009time}. The vertical shift factor $b_T$ is shown in Fig. \ref{Figure:7}c.

\section{\label{sec:level4}Summary and Conclusion}
	We studied the viscoelastic and transport behaviors of poly(tetramethylene oxide) melts, using molecular dynamics simulations of the united-atom model polymer chains. For fully equilibrated polymer chains, we analyzed their mean-squared-displacement and discussed our findings in context with the recent experimental observations. By considering the full chain and only the middle portion, we verified the Rouse and reptation theory predictions. In context with the experimental observations, this study underlines that, resolving the position-dependent displacement with accurate scaling at intermediate timescale, requires that different chain portions are considered along the chain length. Given that the modeled PTMO chains are semi-flexible with comprehensive atomistic detail it is challenging to resolve all the scaling regimes observed experimentally \cite{abadi2018entangled}.

	From our extensive molecular dynamics simulations, we have probed the time-dependent relaxation behavior of the PTMO chains for a range of molecular weights and temperatures. The shear relaxation modulus that describes the viscoelasticity of a polymer melt is investigated along with the bulk relaxation modulus. The shear relaxation modulus showed the Rouse behavior in $G(t)$ with scaling law $t^{-1/2}$. At long times, the relaxation moduli reaches a plateau value and as the chain size increases this plateau extends along the time scale as predicted by reptation theory. We have also studied the bulk relaxation modulus for a range of molecular weights and it is found to be dependent on the chain length, where the relaxation time for $K(t)$ increases with increasing molecular weight.

	Finally, we determined the master curves of both the shear and bulk viscoelastic behavior of the PTMO by using the time-temperature superposition principle. To construct master curves, the calculated horizontal shift factor is found to be consistent with the WLF equation and with previous work \cite{rouleau2013application}. 

	The obtained time-dependent viscoelastic properties of PTMO can be used as an input for meso- or macroscale simulations of morphology and rheology of copolymers and aid the design of polymeric materials. The outcomes of this paper can also be used to compute the frequency dependent relaxation data for continuum-level computational studies of a multiscale material system with PU as the matrix. Since the results are directly associated with molecular features, they can provide deeper insights to better tune the microstructures of PU and enhance the materials performance.

\begin{acknowledgements}
	This research is supported by the Commonwealth of Australia as represented by the Defence Science and Technology Group of the Department of Defence. We also acknowledge and thank Spartan-HPC and Argali-HPC at Faculty of Engineering and Information Technology, The University of Melbourne for providing us the computing facility.
\end{acknowledgements}


\providecommand{\noopsort}[1]{}\providecommand{\singleletter}[1]{#1}%

\end{document}